\documentclass[sigconf]{acmart}

\usepackage{tabularx} 

\AtBeginDocument{%
  }

\copyrightyear{2025}
\acmYear{2025}
\setcopyright{rightsretained}
\acmConference[RecSys '25]{Proceedings of the Nineteenth ACM Conference on Recommender Systems}{September 22--26, 2025}{Prague, Czech Republic}
\acmBooktitle{Proceedings of the Nineteenth ACM Conference on Recommender Systems (RecSys '25), September 22--26, 2025, Prague, Czech Republic}\acmDOI{10.1145/3705328.3748113}
\acmISBN{979-8-4007-1364-4/2025/09}

\acmDOI{10.1145/3705328.3748113}

\begin{document}

\title{Item-centric Exploration for Cold Start Problem}

\author{Dong Wang}
\affiliation{%
  \institution{Google LLC}
  \city{Mountain View}
  \state{California}
  \country{USA}
}
\email{wangdo@google.com}

\author{Junyi Jiao}
\affiliation{%
  \institution{Google LLC}
  \city{Mountain View}
  \state{California}
  \country{USA}
}
\email{junyijiao@google.com}

\author{Arnab Bhadury}
\affiliation{%
  \institution{Google LLC}
  \city{Vancouver}
  \state{British Columbia}
  \country{Canada}
}
\email{arniebh@google.com}

\author{Yaping Zhang}
\affiliation{%
  \institution{Google LLC}
  \city{Mountain View}
  \state{California}
  \country{USA}
}
\email{yapingzhang@google.com}

\author{Mingyan Gao}
\affiliation{%
  \institution{Google LLC}
  \city{Mountain View}
  \state{California}
  \country{USA}
}
\email{mingyan@google.com}

\author{Onkar Dalal}
\affiliation{%
  \institution{Google LLC}
  \city{Mountain View}
  \state{California}
  \country{USA}
}
\email{onkardalal@google.com}
\renewcommand{\shortauthors}{Wang et al.}

\begin{abstract}
Recommender systems face a critical challenge in the item cold-start problem, which limits content diversity and exacerbates popularity bias by struggling to recommend new items. While existing solutions often rely on auxiliary data, but this paper illuminates a distinct, yet equally pressing, issue stemming from the inherent user-centricity of many recommender systems. We argue that in environments with large and rapidly expanding item inventories, the traditional focus on finding the "best item for a user" can inadvertently obscure the ideal audience for nascent content. To counter this, we introduce the concept of item-centric recommendations, shifting the paradigm to identify the optimal users for new items. Our initial realization of this vision involves an item-centric control integrated into an exploration system. This control employs a Bayesian model with Beta distributions to assess candidate items based on a predicted balance between user satisfaction and the item's inherent quality. Empirical online evaluations reveal that this straightforward control markedly improves cold-start targeting efficacy, enhances user satisfaction with newly explored content, and significantly increases overall exploration efficiency.
\end{abstract}



\maketitle

\section{Introduction}

Recommending new items to users, a challenge commonly known as the item cold-start problem, is a critical hurdle for modern recommendation systems. New items, due to their inherent lack of sufficient interaction data, are difficult to effectively expose to relevant users. This limitation not only impedes the discovery of novel content but also exacerbates popularity bias, where already popular items receive disproportionate recommendations, thereby hindering content diversity. Given the dynamic nature of recommendation systems and the continuous influx of new items, effectively addressing the item cold-start problem is paramount. It empowers content creators and small businesses by facilitating the identification of initial audiences for their offerings and provides substantial user utility by continuously enriching the recommendable item corpus \cite{Su2024}.

Extensive research has sought to mitigate the item cold-start problem by tackling the challenge of limited interactions. Common approaches include leveraging auxiliary data such as item attributes or side information \cite{Wang2019,Chen2024,Xie2022} or employing advanced learning paradigms designed for data scarcity, such as meta-learning or transfer learning \cite{Volkovs2017,Wei2023,Huang2022,Vartak2017,Xu2022}. However, in this paper, we highlight a distinct and often overlooked challenge: even with existing cold-start mechanisms applied, a purely user-centric approach can lead to suboptimal initial audience targeting for new items. This suboptimality can significantly impede an item's subsequent broader distribution by the main exploitation system, a problem that is particularly exacerbated in extremely large item spaces where the volume of new items is immense. Research specifically addressing this latter challenge remains limited \cite{Chen2025}. We argue that, in contrast to the prevailing user-centric paradigm, a more natural and effective approach for item cold-start is an item-centric one: proactively identifying the most suitable users for a given new item.

\begin{figure}[h]
  \centering
  \includegraphics[width=\linewidth]{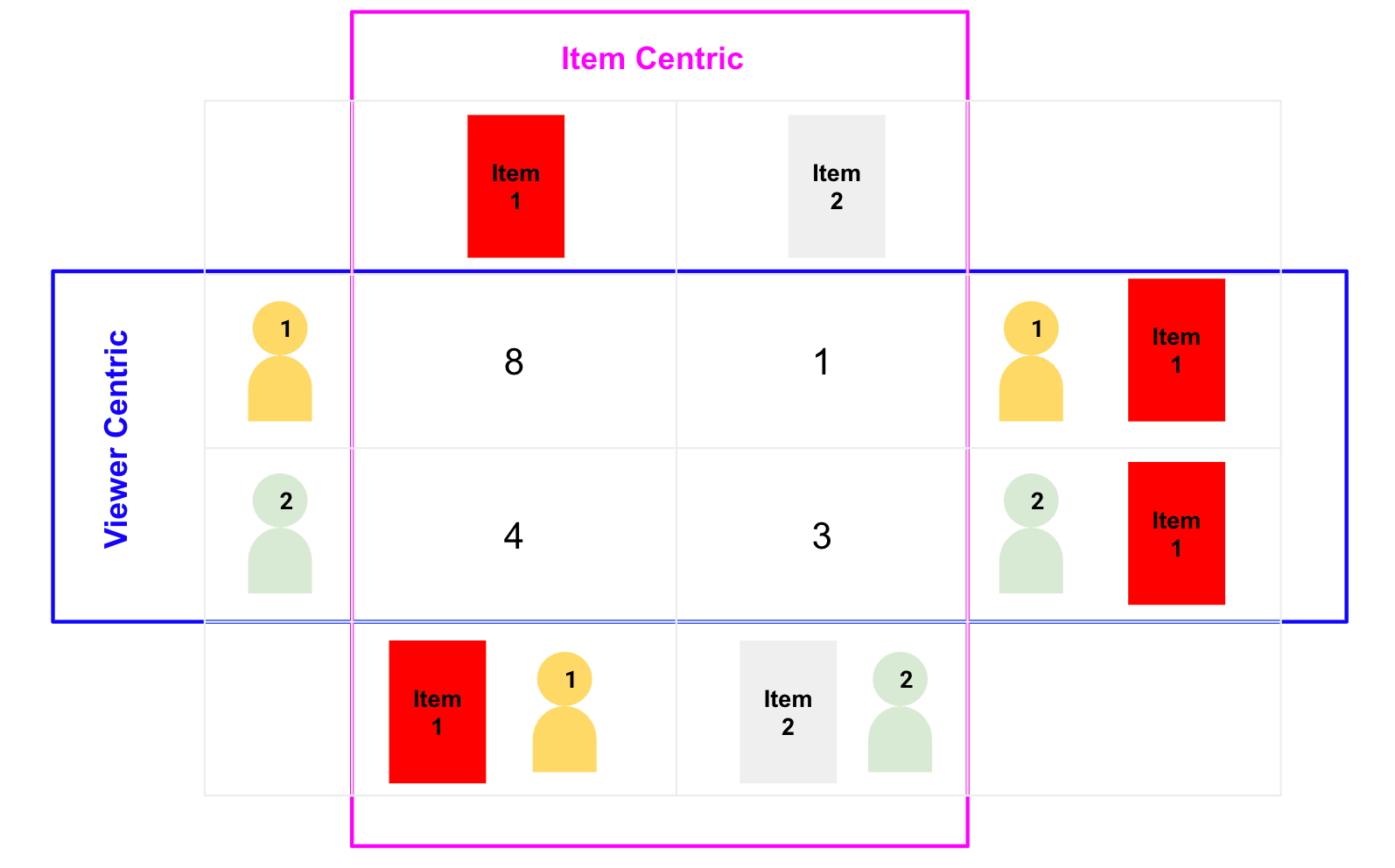}
  \caption{Illustration the asymmetry between user-centric and item-centric recommendations. A 2x2 table shows user-item positive scores. Blue highlights user-centric recommendations (best items for a user); purple highlights item-centric recommendations (best users for an item).}
  \label{fig:exploration_matrix}
  \Description{The figure illustrates a 2x2 matrix of user-item scores. Blue circles show how user-centric recommendations pick the best item per user (horizontally). Purple circles show how item-centric recommendations pick the best user per item (vertically), highlighting the conceptual asymmetry.}
\end{figure}

We contend that an item-centric approach is better equipped to help long-tail items discover their initial audiences, while simultaneously improving the initial targeting quality for even potentially popular items. To further illustrate this crucial distinction, Figure \ref{fig:exploration_matrix} provides a toy example. It demonstrates how, within a user-centric system, a popular item might be inadvertently matched with a suboptimal audience, and critically, how a long-tail item might fail to find its optimal audience altogether, thus never gaining traction.

As a significant step toward building item-centric exploration, we implemented and evaluated an item-centric control within our existing exploration system. Details regarding the implementation of this item-centric control are provided in Section \ref{method}. We experimented with this new component on real traffic within a large-scale short video recommendation system, and the results of these experiments are presented in Section \ref{results}.

\section{Methodology}
\label{method}
\subsection{Overview}
In our exploration system, when a user submits a request, a set of relevant candidate items is initially retrieved from a pool of new content based on the user's inferred interests. These candidates are then ordered by a ranking system before being presented to the user. To specifically promote items that demonstrate the potential for high satisfaction within a targeted audience, we introduce a novel item-centric filtering component after the ranking stage. This control strategically evaluates each item before its presentation to the user.

\begin{figure}[h]
  \centering
  \includegraphics[width=0.8\linewidth]{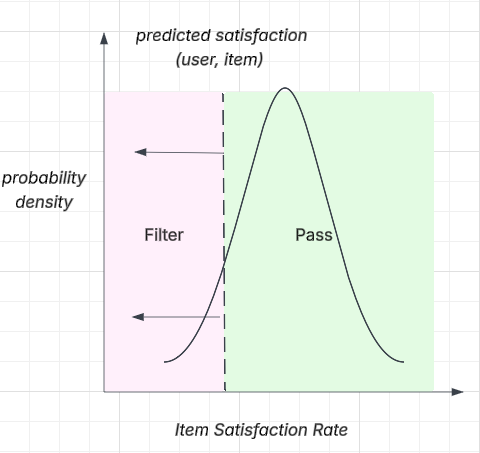}
  \caption{Posterior Distribution of the Item's Satisfaction Rate. If the predicted satisfaction of the user is significantly less than the satisfaction rate posterior mean, we then remove the item from the candidate list.}
  \label{fig:overview}
  \Description{An item is filtered if its predicted user-item satisfaction probability is more than two standard deviations below its average satisfaction rate.}
\end{figure}

The core principle of this filter is to avoid exposing an item to a user if the predicted user's satisfaction for that item is significantly lower than the item's inherent quality, as indicated by its historical average satisfaction rate. As depicted in Figure \ref{fig:overview}, when the predicted user-item satisfaction probability falls considerably below the item's average satisfaction rate, this suggests that the current audience is likely a mismatch for the item, making its display to this "wrong" audience inefficient. The satisfaction rate for an item is calculated by dividing the total count of satisfactions it has received by its total number of impressions.

Mathematically, the primary condition for an item to be filtered is:
\begin{equation}
  p(\text{S}_{+} | \text{u}, \text{i}) < \mu_{\text{i}} - 2 \sigma_{\text{i}},
\end{equation}
where $\text{S}_+$ represents a satisfied interaction, $\mu_{\text{i}}$ and $\sigma_{\text{i}}$ represent the posterior mean and posterior standard deviation, respectively, of the item's satisfaction rate.

\subsection{Predicting User's Satisfaction Probability}
The probability of a satisfied impression $p(\text{S}_+ | \text{u}, \text{i})$ is obtained from the output of a large multi-task ranking model with various prediction heads \cite{Zhao2019}.

Given that the filter relies on the output of the ranking stage, specifically the prediction heads, it is necessarily positioned after the ranking stage.

\subsection{Modeling the Item's Intrinsic Satisfaction Rate}
We model the intrinsic satisfaction rate of an item using a Beta distribution. This choice is motivated by the fact that the Beta distribution is a conjugate prior for the Bernoulli distribution, which naturally models binary outcomes like satisfied or not satisfied. The conjugacy allows the model to be extremely memory efficient and only store a couple of parameters to know the full distribution. The posterior mean and variance can be calculated on the fly. As an item accumulates more impression data, the observed satisfaction rate converges towards its true, underlying intrinsic rate.

We define the prior distribution for the satisfaction rate as a Beta distribution with parameters $\alpha_0$ and $\beta_0$, denoted as $B(\alpha_0, \beta_0)$. Given satisfactions count $N_+$ and total impressions $N$ for an item, the posterior distribution of its satisfaction rate is also a Beta distribution, specifically $B(\alpha_0 + N_+, \beta_0 + N - N_+)$.

The posterior mean ($\mu_{\text{i}}$) and variance ($\sigma^2_{\text{i}}$) of the satisfaction rate can then be calculated as:
\begin{equation}
  \mu_{\text{i}} = \frac{\alpha_0 + N_+}{\alpha_0 + \beta_0 + N},
\end{equation}
and
\begin{equation}
  \sigma^2_{\text{i}} = \frac{(\alpha_0 + N_+)(\beta_0 + N - N_+)}{(\alpha_0 + \beta_0 + N)^2 (\alpha_0 + \beta_0 + N + 1)}.
\end{equation}

\begin{figure}[h]
  \centering
  \includegraphics[width=0.8\linewidth]{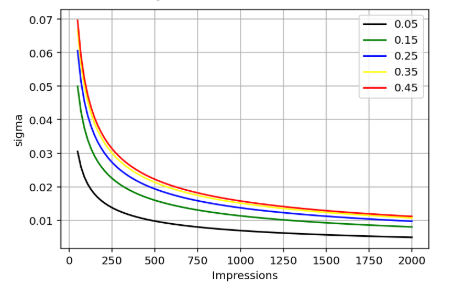}
  \caption{Convergence of Standard Deviation of satisfaction Rate.}
  \label{fig:sigma_playbacks}
  \Description{The standard deviation of the estimated satisfaction rate decreases as the total number of impressions increases. The rate of convergence is also influenced by the underlying satisfaction rate.}
\end{figure}

The standard deviation ($\sigma_{\text{i}}$), which is the square root of the variance, decreases as the item accumulates more impressions, indicating increasing confidence in the estimated satisfaction rate. Figure \ref{fig:sigma_playbacks} illustrates this convergence of $\sigma_{\text{i}}$ as a function of the total number of impressions for different underlying satisfaction rates.

As the distribution of the satisfaction rate evolves rapidly during the initial few hundred impressions, it is crucial that the updates to the posterior mean ($\mu_{\text{item}}$) and standard deviation ($\sigma_{\text{item}}$) are performed with low latency. This necessitates the low-latency aggregation of user interaction statistics.

\section{Results}
\label{results}
\begin{figure}[h]
  \centering
  \includegraphics[width=0.8\linewidth]{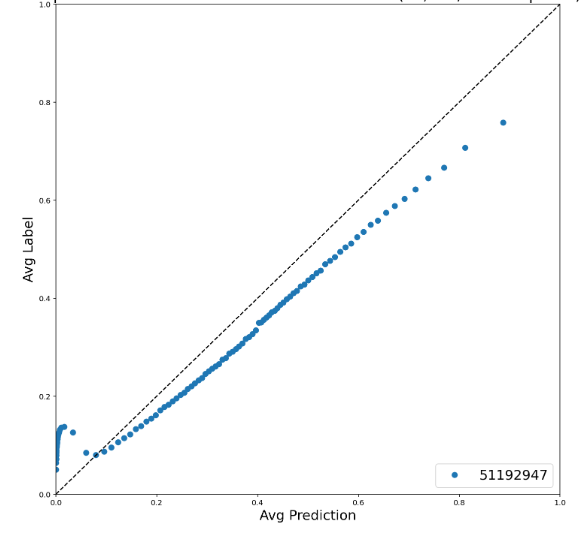}
  \caption{Calibration of Satisfaction Predictions.}
  \label{fig:calibration}
  \Description{The model tends to underestimate the satisfaction rate for items with very low observed rates and slightly overestimate it for items across other ranges.}
\end{figure}

To assess the accuracy and reliability of our model's satisfaction predictions for (user, item) pairs, we conducted a calibration analysis. Figure \ref{fig:calibration} presents a calibration plot that compares the average predicted satisfaction rates within discrete prediction bins against the corresponding observed average satisfaction rates (ground truth) for items falling into those bins.

Overall, the model's predictions demonstrate good alignment with the ground truth. While a degree of miscalibration is observed in the region of very low satisfaction rates, empirical analysis indicates that this minor deviation does not detrimentally affect the efficiency or performance of the proposed item-centric control.

\subsection{Online Experiment Results}
The exploration system aims to expand the recommendable item corpus while mitigating potential negative impacts on short-term user satisfaction \cite{Chen2021}. While user satisfaction can be effectively measured through user-diverted experiments, evaluating the expansion of the recommendable corpus necessitates a distinct corpus-level measurement. This is because traditional user-diverted experiments cannot apply distinct treatments to disjoint item sets, which is essential for directly assessing changes in the item corpus. To address this, we developed and implemented a user-corpus co-diverted exploration experiment framework. This design enables the simultaneous evaluation of both user satisfaction metrics and the expansion of the recommendable corpus.

The impact of this change on key user satisfaction metrics, as observed in our live experiments, is presented in Table \ref{tab:live-exp-result}.

\begin{table}
  \caption{Live Experiment Result}
  \label{tab:live-exp-result}
  \begin{tabularx}{\linewidth}{l|X|X|X|X} 
    \toprule
    & Satisfaction Metric 1 & Satisfaction Metric 2 & Exploration Impression & Recommendable Corpus \\
    \midrule
    Treatment & +50\% & +40\% & -20\% & +10\% \\
    \bottomrule
  \end{tabularx}
  \Description{The treatment has significantly improved both user satisfaction metrics and recommendable corpus size.}
\end{table}

The results demonstrate significant positive gains across user satisfaction metrics. As anticipated, the introduction of the filtering mechanism led to a 20\% reduction in overall exploration item impressions, as the system is more selective when finding audiences for items. Crucially, however, the satisfaction of users with the explored items significantly improved, indicating an overall enhancement in the quality of exploration items presented.

Furthermore, we observed a remarkable 10\% increase in the size of the recommendable corpus, which is a substantial efficiency improvement of the whole system and represents a significant breakthrough. Importantly, this expansion was achieved alongside a reduction in exploration load (fewer impressions) and a concurrent improvement in user satisfaction with the explored content, validating the effectiveness of our exploration system in achieving its dual objectives.

\section{Conclusion}
We introduced an item-centric approach to efficiently address item cold-start problems by identifying the most suitable audience for each item. Our working solution significantly improved exploration efficiency, quality, and user satisfaction. This lightweight item-centric control, more selective in finding audiences, greatly improved cold-start targeting quality while avoiding significant operational risk compared to a complete system transition.

\section{Speaker Bio}
Dong Wang is a Software Engineer at Google (YouTube). His work primarily
focuses on short-form video recommendations.


\bibliographystyle{ACM-Reference-Format}
\bibliography{exploration-base}

\appendix

\end{document}